\documentclass[aps,twocolumn,showpacs,showkeys,groupedaddress,amsmath,amssymb]{revtex4}

\usepackage{graphicx}
\usepackage{dcolumn}
\usepackage{bm}

\begin{document}


\title{Computer simulation of coherent interaction of charged particles and
photons with crystalline solids at high energies}


\author{Armen Apyan}
\email[E-mail address:\,] {aapyan@lotus.phys.northwestern.edu}

\affiliation{Northwestern University, Department of Physics and Astronomy, 2145
Sheridan Road, Evanston, IL 60208, USA}



\begin{abstract}
Monte Carlo simulation code has been developed and tested for studying the
passage of charged particle beams and radiation through the crystalline matter
at energies from tens of MeV up to hundreds of GeV. The developed Monte
Carlo code simulates electron, positron and photon shower in single crystals
and amorphous media. The Monte Carlo code tracks all the generations of
charged particles and photons through the aligned crystal by taking into
account the parameters of incoming beam, multiple scattering, energy loss,
emission angles, transverse dimension of beams, and linear polarization of
produced photons.
                                                                                
The simulation results are compared with the CERN-NA-59 experimental data. The
realistic descriptions of the electron and photon beams and the physical
processes within the silicon and germanium single crystals have been
implemented.
\end{abstract}


\maketitle


\section{Introduction}
\label{intro}

In the last decade, the electromagnetic interaction of charged particles and
photons with crystalline and amorphous media are intensively investigated both
theoretically and experimentally. Special attention was given to the
investigation of interaction of charged particles and photons with
crystalline media. An electron or photon impinging on a crystal will interact
coherently with the atoms in aligned crystal axes or planes. If the Laue
condition is satisfied, the coherent bremsstrahlung~(CB) or coherent pair
production~(CPP) phenomena are manifested~\cite{ter-misha}. The
essential characteristics of the phenomena are quasi-monochromatic spectrum,
high intensity and linear polarization degree of radiation in coherent maximum.
The intensity of the coherent bremsstrahlung in aligned crystals is few ten
times greater than in amorphous media at the certain energy region. The same
characteristics are seen in CPP by photons in the single crystals. The
processes of CB and CPP in single crystals are well investigated and
understood both theoretically~\cite{ter-misha} and
experimentally~\cite{apyan1,apyan2}.

In this work we describe briefly the physical processes involved, the
simulation model and the results of the Monte Carlo simulations for high
energy particles traversing an aligned single crystals.
                                                                                
It is well known that the dominant energy loss mechanism for high energy
electrons and positrons is the production of electromagnetic radiation, i.e.
bremsstrahlung for motion through the matter. The high energy photons become
absorbed mainly due to the $e^{+}e^{-}$ pair production in a matter. This is
true for crystals as well. In our simulations we ignored the processes such as
Compton scattering, photoeffect, energy loss in ionization, nuclear processes
due to smallness of their cross section and negligible contribution in total
cross section. Electron or photon beams penetrating the crystal will create an
electromagnetic shower~(EMS) via the CB, CPP or incoherent bremsstrahlung~(ICB)
and pair production~(IPP) depending on the orientation of crystal axes and
planes relative to incident particle momentum.

There are broad experimental and theoretical investigations devoted to the
EMS development in various amorphous media and energies. Many
general - purpose Monte Carlo simulation packages (see Ref.~\cite{jenkins}
and references cited therein) for transport of particles and radiation through
the amorphous media are developed and have been successfully used in the last
decades.
                                                                                
The electromagnetic interaction processes in aligned crystals are more
complicated than in amorphous media. The cross section of interaction strongly
depends on the crystal type, its orientations as well as on the energy,
angular distribution and polarization of initial photon or electron beams.
                                                                                
The EMS in single crystals was mainly considered in  channeling regime, when
electrons and photons penetrate the crystal along a direction close to the one
of main crystallographic axes~\cite{apyan-shower,baskov}. We consider EMS in
single crystals oriented in CB mode. All the above mentioned peculiarities of
coherent processes are carefully taken into account in Monte Carlo computer
code.

\section{Theoretical background}
\label{theory}
                                                                                
Differential cross section of coherent radiation in a crystal is composed of
two terms~\cite{ter-misha} coherent and incoherent bremsstrahlung
\begin{equation}
d \sigma = d \sigma_{inc} + d \sigma_{coh}
\label{eq:sigma}
\end{equation}

The first term corresponds to the incoherent cross section (including
radiation in the field of the atomic nucleus and electrons) on $N$ independent
atoms. The second one corresponds to the coherent radiation cross section.
Let us denote ($E_0, \mathbf P_0$), ($E_1, \mathbf P_1$) and ($E_{\gamma},
\mathbf P_{\gamma}$) as the energy and momentum of incoming, outgoing electrons
and emitted photon, respectively. The initial beam orientation with respect
to the crystal axes is defined by two angles in the following manner. The
three chosen orthogonal axes of cubic crystal are
$[\mathbf {b_1 \, b_2 \, b_3}]$. The initial beam orientation is defined by
the angle $\theta_0$ between the initial electron momentum $\mathbf P_0$ and
crystal axis $\mathbf b_3$ and by the angle $\psi_0$ between the electron
momentum and crystal plane $(\mathbf {b_1 \, b_3})$. Let $\theta$ and
$\varphi$ be the emitted photon polar and azimuthal angles with respect to the
direction of initial motion of the electron. Usually the polar angle is
presented in the units of $m c^2/E_{in}$:
\begin{equation}
u = \frac{E_{in}}{m c^2} \theta
\label{eq:u-angle}
\end{equation}
where $E_{in}$ is the energy of initial particle, $m$ is the electron rest mass
and $c$ is speed of light.
                                                                                
The angular-spectral distribution of CB (after integration with respect to
exit angles of electrons) is given by the following
expressions~\cite{ter-misha}:
\begin{widetext}
\begin{equation}
d \sigma^3 (\mathrm x, \theta_0, \alpha_0, \xi, \varphi) =
\frac {N \sigma_0} {2 \pi} \, \frac {d \mathrm x} {\mathrm x} \, d \xi \,
d \varphi \, I (\mathrm x, \theta_0, \alpha_0, \xi, \varphi)
\label{eq:dif-cross}
\end{equation}
\begin{equation}
I \big(\mathrm x, \theta_0, \alpha_0, \xi, \varphi \big) =
\Big[ 1 + (1 - x )^2 \Big] \Big(\Psi^{coh}_1 +
\Psi^{inc}_1 + \Psi^{el}_1 \Big)  - 
\frac{2} {3} \Big( 1 - \mathrm x \Big) \Big(\Psi^{coh}_2 + \Psi^{inc}_2 +
\Psi^{el}_2 \Big)
\label{eq:int}
\end{equation}
\end{widetext}
where $ I (\mathrm x, \theta_0, \alpha_0, \xi, \varphi)$ is the intensity of
radiation, $N$ is the number of atoms in a crystal, 
$\sigma_0 = Z^2 r_0^2 \alpha$, $Z$ is the atomic number of medium, $r_0$ is 
the classical electron radius, $\alpha$ is the fine structure constant, 
$\mathrm x = E_{\gamma} / E_0$ is the relative energy of emitted photon and 
$\xi = 1/(1+u^2)$.

The $\psi^{coh}_{1,2}$ functions in equation Eq.~\ref{eq:int} have the
following structure~\cite{ter-misha}:
\begin{eqnarray}
\Psi^{coh}_1 &=& 4 \sum_{\mathbf g} D_g \, g_{\perp}^2
\delta_D \biggl[ \frac {\delta}{\xi} - g_{\parallel} \biggr]
\nonumber \\
\Psi^{coh}_2 &=& 24 \sum_{\mathbf g} D_g \, \xi ( 1 - \xi )
g_{\perp}^2 \delta_D \biggl[ \frac {\delta}{\xi} - g_{\parallel} \biggr]
\nonumber \\
D_g &=& \frac{( 2 \pi )^2 } {\Delta}
\frac { \mid S(\mathbf g ) \mid ^2} {N_0}
\frac {\big[ 1 - F(g) \big]^2} {g^4} \exp (- \bar{A} g^2)
\label{eq:psi-coh}
\end{eqnarray}
where $\delta_D$ is Dirac's delta function, $\mathbf g$ is the reciprocal
lattice vector, $\Delta$ is the volume of an elementary cell of a direct
lattice, $S(\mathbf g)$ is the structure factor of the crystal, $\bar A$
is the root mean square of thermal displacement amplitude of an atom from an 
equilibrium position. In our calculations we used Doyle-Turner
parameterization~\cite{DT} for the atomic formfactor $F(g)$. For a given
orientation of the crystal, the longitudinal $g_{\parallel}$ and transverse
$g_{\perp}$ component of $\mathbf g$ with respect to the initial electron
momentum $\mathbf P_0$ are:
\begin{eqnarray}
g_{\parallel} &=& g_3 \cos \theta_0 + (g_1 \cos \alpha_0 + g_2 \sin \alpha_0)
\sin \theta_0
\nonumber \\
g_{\perp}^2 &=& g^2 - g_{\parallel}^2
\label{eq:g-vector}
\end{eqnarray}
where $g_1, g_2, g_3$ are the projections of $\mathbf g$ on the crystal axes
$[b_1 \,b_2 \,b_3]$. The $\hbar \delta$ is a minimal value of the momentum
transferred to medium along the direction of motion of primary particle:
\begin{equation}
\hbar \delta = \frac{\hbar E_{\gamma} m c^2}{2 E_0 E_1} m c
\label{eq:delta}
\end{equation}
where $\hbar$ is the reduced Planck's constant.

The incoherent $\psi^{inc}_{1,2}$ functions in equation Eq.~\ref{eq:int} have
the following structure~\cite{olsen}:
\begin{eqnarray}
\Psi^{inc}_1 &=& 6 + 4 \Gamma(\xi)
\nonumber \\
\Psi^{inc}_2 &=&  6 + 24 \xi (1 - \xi) \Gamma(\xi)
\label{eq:psi-inc}
\end{eqnarray}
where
\begin{equation}
\Gamma(\xi) = \ln \bigg( \frac{m c}{\hbar \delta} \bigg) -2 -
f \Big(\zeta \Big) + \mathfrak{F} \big(\hbar \delta/ \xi \big)
\end{equation}
is the general expression for $\Gamma \big(\xi \big)$ for arbitrary screening.
The quantity $\mathfrak{F} \big(\hbar \delta/ \xi \big)$ has the following
form~\cite{olsen}:
\begin{widetext}
\begin{equation}
\mathfrak{F} \big(\hbar \delta/ \xi \big) = \int_{\hbar \delta / \xi}^{\infty}
\Big\{ 1 - \exp \Big(- \frac {\bar{A} q^2} {\hbar^2} \Big) \Big\}
\Big\{ \big[1-F(q)^2 \big] -1 \Big\}
\frac{(q^2 - (\hbar \delta)^2 / \xi^2)}{q^3} dq
\label{eq:F-dzevavor}
\end{equation}
\end{widetext}
where $q$ is the momentum transferred to the nucleus. The quantity
$f \big(\zeta \big)$ is the Coulomb correction:
\begin{equation}
f \big(\zeta \big) = \zeta^2 \sum_{n=1}^{\infty}
\frac{1}{n \big(n^2 + \zeta^2 \big)}
\label{eq:coulomb-corr}
\end{equation}
where $\zeta = Z \alpha$. The corrections due to contributions by the atomic
electrons, i.e. quantities $\Psi^{el}_{1,2}$ were calculated by the theory
given in~\cite{lamb}.

The threefold differential cross section of CPP by photons in single crystals
can be written in the following form~\cite{ter-misha}:
\begin{widetext}
\begin{equation}
d \sigma^3 (\mathrm y, \theta_0, \alpha_0, \xi, \varphi) =
\frac {N \sigma_0} {2 \pi} \, d \mathrm y \, d \xi \, d \varphi \,
I (\mathrm y, \theta_0, \alpha_0, \xi, \varphi)
\label{eq:dif-pp}
\end{equation}
\begin{equation}
I \big(\mathrm y, \theta_0, \alpha_0, \xi, \varphi \big) =
\Big[ \mathrm y^2 + (1 - \mathrm y )^2 \Big] \Big(\Psi^{coh}_1 + \Psi^{inc}_1 +
\Psi^{el}_1 \Big) + 
\frac{2} {3} \, \mathrm y \Big( 1 - \mathrm y \Big) \Big(\Psi^{coh}_2 +
\Psi^{inc}_2 + \Psi^{el}_2 \Big)
\label{eq:int-pp}
\end{equation}
\end{widetext}

Here we use the following notation. The ($E_{\gamma}, \mathbf P_{\gamma}$) are
the initial photon energy and momentum and ($E_{+}, \mathbf P_{+}$),
($E_{-}, \mathbf P_{-}$) are the energy and momentum of produced positron and
electron pair. The $\mathrm y = E_{+} / E_{\gamma}$ is the relative energy of
produced positron, $u = \big( E_{+} / m c^2) \theta_{+} $ and $\theta_{+}$ is
the angle between the positron and photon momenta. In case of CPP a minimal 
value of the momentum transferred to medium along the direction of motion of 
primary particle has the form:
\begin{equation}
\hbar \delta = \frac{\hbar E_{\gamma} m c^2}{2 E_{+} E_{-}} m c
\label{eq:delta-pp}
\end{equation}
                                                                                
The above described parameters are the same as for CB theory.

\begin{figure}[htbp]
\includegraphics[scale=0.355]{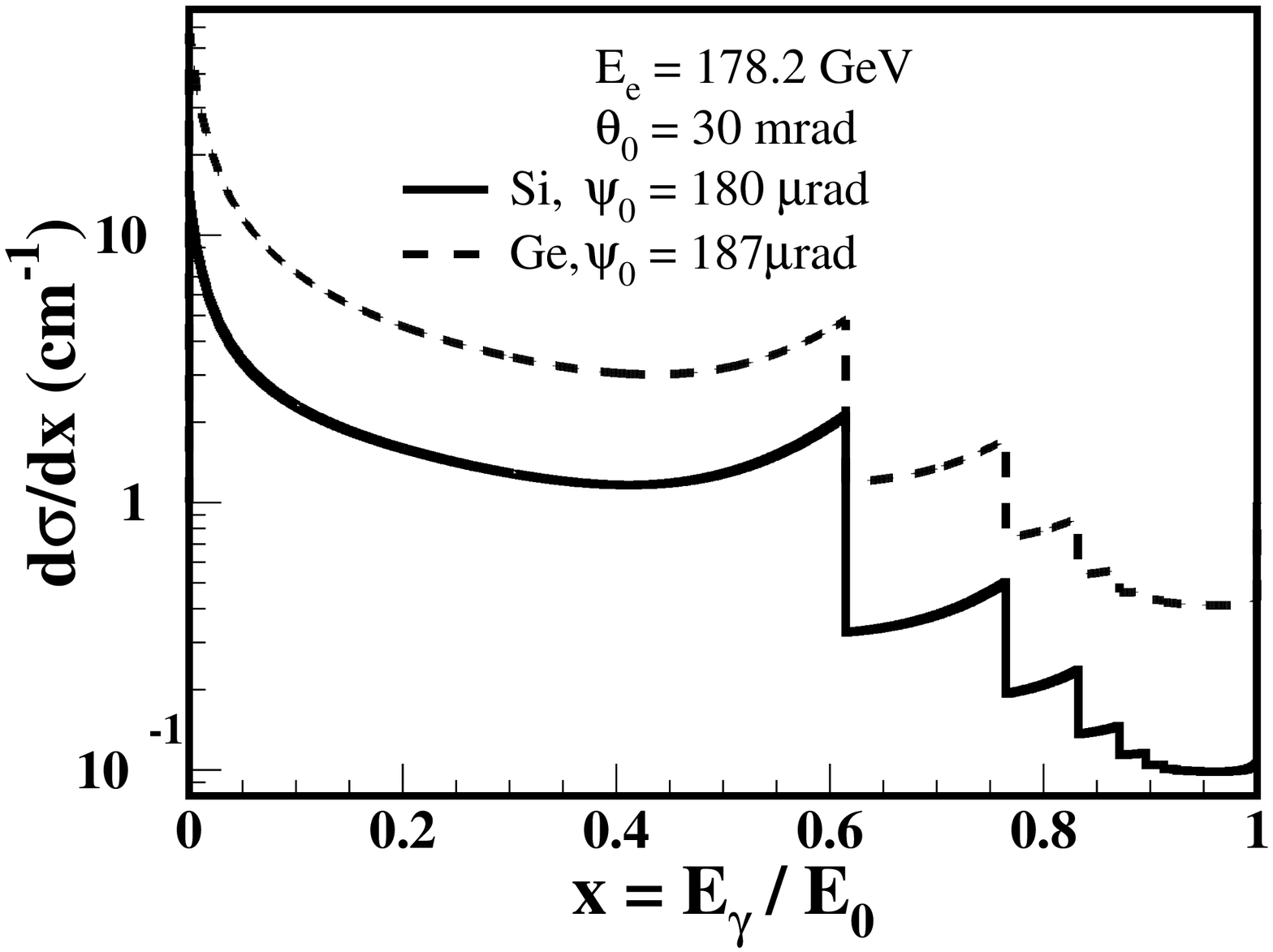} \\
\includegraphics[scale=0.355]{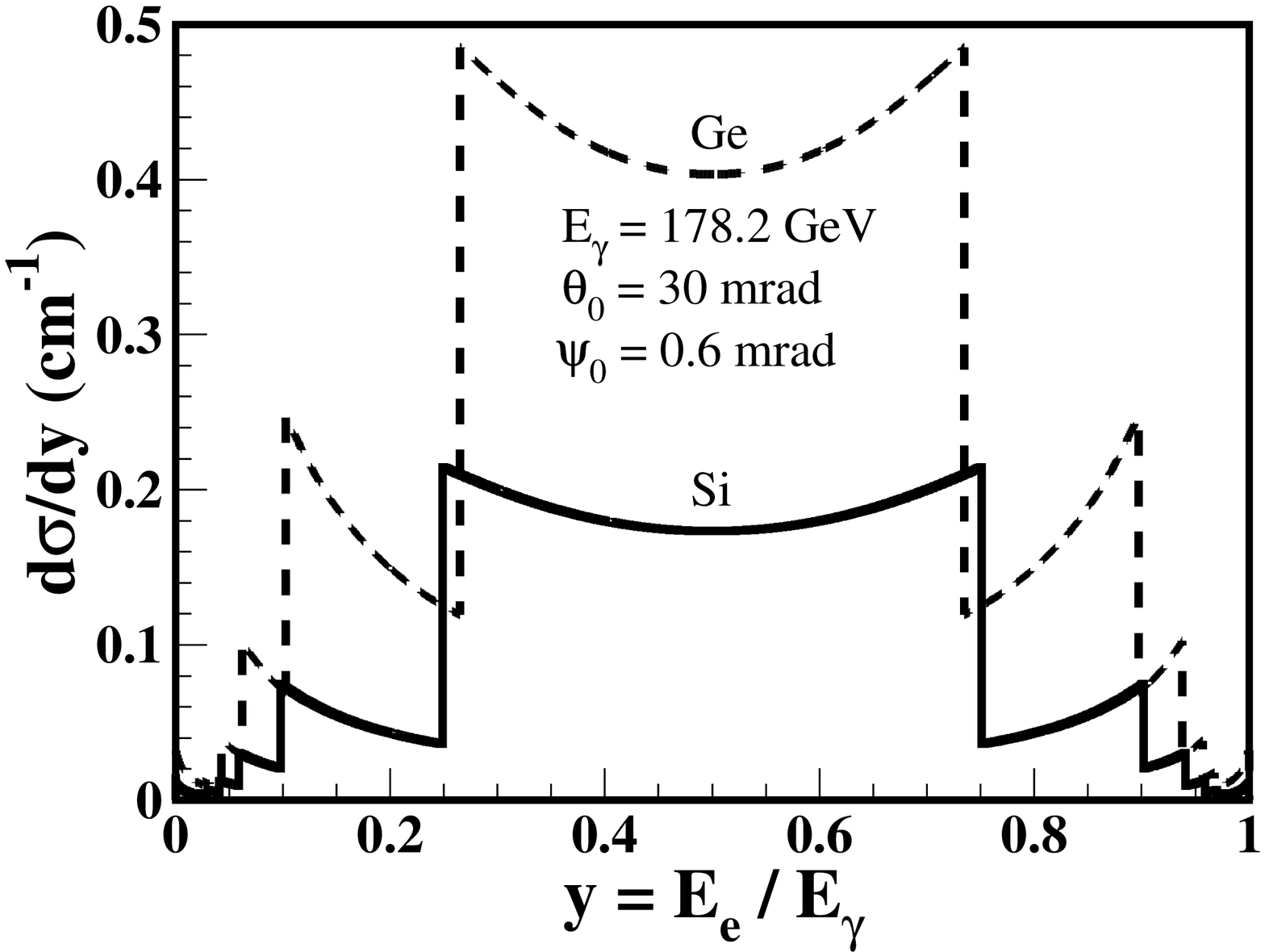}
\caption{\label{F:cros-sec} Differential cross sections of CB~(top) as a
function of photon energy, $\mathrm x = E_{\gamma} / E_0$, and CPP~(bottom) as
a function of the energy of the $e^{+}e^{-}$ pair component,
$\mathrm y = E_{\pm} / E_{\gamma}$ for Si~(solid curves) and
Ge~(dashed curves) single crystals.}
\end{figure}

Computed differential cross sections of CB~(Eq.~\ref{eq:dif-cross}
integrated over the emitted photon angles ($\theta$, $\varphi$)) and
CPP~(Eq.~\ref{eq:dif-pp} integrated over the production angles
($\theta_{\pm}$, $\varphi_{\pm}$) of $e^{+}e^{-}$ pairs) are given in
Fig.~\ref{F:cros-sec}.

The top figure represents the spectral distribution of CB depending on the 
photon energy, $\mathrm x = E_{\gamma} / E_0$, on the Si and Ge single 
crystals. For an $E_0 = 178.2 GeV$ electron beam making an angle of 
$\theta_0 = 30~mrad$ from the $<$001$>$ crystallographic axis and 
$\psi_0 = 180 \mu rad$ from the $(110)$ plane of Si crystal, the maximum peak 
intensity occurs in the vicinity of 100~GeV as seen in Fig.~\ref{F:cros-sec}. 
The angle $\psi_0 = 187 \mu rad$ was chosen for the corresponding spectral 
distribution in case of Ge crystal.
                                                                                
The differential cross sections of CPP in Si and Ge single crystals are given
in the bottom of Fig.~\ref{F:cros-sec} depending on the energy of the
$e^{+}e^{-}$ pair component, $\mathrm y = E_{\pm} / E_{\gamma}$. The total
cross section of CPP has a maximum value for the chosen primary photon energy
$E_\gamma = 178.2 GeV$ and orientation angles $\theta_0 = 30~mrad$ from the
$<$001$>$ crystallographic axis and $\psi_0 = 0.6 mrad$ from the $(110)$
plane.

Particles could undergo transmission at small angles with respect to the 
crystal axes and planes due to angular divergence and multiple scattering 
during the EMS development in crystal. CB and CPP processes have
strong angular and energy dependences and the validity conditions of the
Born approximation no longer hold at very high energies and small incidence
angles with respect to the crystal axes and planes. The onset of this problem
for the description of radiation emission and pair production has the
characteristic angle $\theta_v=U_0/m c^2$~\cite{baier}, where $U_0$ is the
plane potential well depth. The CB and CPP theory may be applied for the
incidence angles with respect to the crystal axes/planes 
$\theta_0 \gg \theta_v$. The general theory of radiation and pair production 
is used~\cite{baier,simon} for incident angles $\theta_0 \sim \theta_v$ and
$\theta_0 < \theta_v$.

\section{Monte Carlo Simulations}
\label{mc}
                                                                                
\subsection{General Considerations}
\label{gen-mc}
                                                                                
Monte Carlo simulation technique of EMS development in oriented single crystals
is based on the coherent radiation and coherent $e^{+}e^{-}$ pair production
processes at high energies. The program includes the formulae of coherent and
quasiclassical theories for radiation and  pair production. The program
calculated the cross sections directly using the
formulae of coherent effects~(Eqs.~\ref{eq:dif-cross} and~\ref{eq:dif-pp}).
The quasiclassiacl theory was applied for calculating the cross sections in
case of small incident angles with respect to the crystal axes or planes.
Direct calculations by the formulae of this theory takes large computer time.
Because the computer time needed to perform all the simulations is formidable,
we used a data bank of pre-calculated cross sections of radiation and pair
production to save the computer time. The data bank contains the total and
differential cross sections on dependence of crystal type and its orientation,
energy, horizontal and vertical angles and polarization. The cross sections
for certain energy, angles, polarization were found by interpolation of the
numerical values of cross sections stored in a data bank. The interpolated
values of cross sections differ from calculated values $\sim 2-3\%$. We find
that our approach yields very accurate numerical results for theory and a 
further saving of computer time.

\subsection{Simulation model}
\label{sim-model}

The simulation model used in this work is based on random walk or particle
history, where a history corresponds to following a particle from entering the
medium, interacting and leaving it. Each particle history $i$ can be
represented by the array $\mathbf S^i_j$~\cite{morin} denoting the state
of particle before $j$th interaction:
\begin{equation}
\mathbf{S}^i_j = \big( \mathbf{r}^i_j , \mathbf{\Omega}^i_j ,
\mbox{\boldmath$\eta$}^i_j , E^i_j , q^i_j \big .)
\label{eq:state}
\end{equation}
where $ \mathbf{r}^i_j , \mathbf{\Omega}^i_j , \mbox{\boldmath$\eta$}^i_j ,
E^i_j , q^i_j $ represent the electron or photon position, direction,
polarization, energy and charge before each interaction act, respectively. 

The initial particles, electrons or photons, generate two secondaries after 
each interaction. The initial electron loses energy via bremsstrahlung and 
produces secondary electron and bremsstrahlung photon. The electron can produce 
several photons during the traveling through matter. The number of produced
photons depend on the thickness of the crystalline target and its orientation.
The initial photon is transformed into the $e^{+}e^{-}$ pair after interaction.
The simulation code calculates new state of particles and state of produced
particles after each interaction act. A history is terminated when the particle
energy drops below a low energy cut-off, or when particle moves outside to the
target.

The similar simplified Monte Carlo simulation model was employed
in~\cite{armen3,armen4}. The simulation procedure has been improved by taking
into account the peculiarities of the emission angles of photons and electrons
in case of coherent effects~(see Eq.~\ref{eq:psi-coh}). This improvement
gives the possibility to use the code for simulation of production of 
collimated photon beams from thin crystals~($\sim 20 - 100 \mu m$)~\cite{e159}.
Polarization dependent cross sections and transverse dimensions of the
particles beams are also implemented in the code.
                                                                                
\section{Simulation results and discussion}
\label{res}
                                                                                
\subsection{Comparison with experiment}
\label{experiment}

A series of Monte Carlo simulations were performed for prediction of the 
results of CERN NA59 experiment. The goal of the experiment was the production 
of linearly polarized photon beams and conversion of the linear polarization 
into circular with the help of oriented single 
crystals~\cite{apyan1,apyan2,apyan3,apyan4,unel}.

\begin{figure}[htbp]
\centering
\includegraphics[scale=.4]{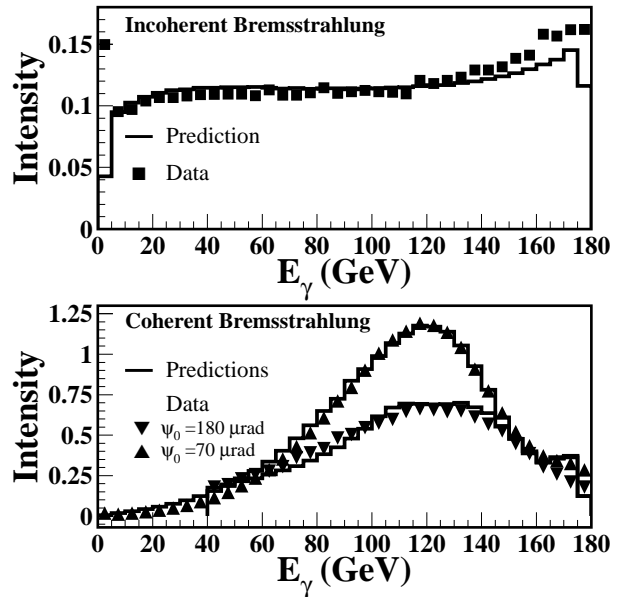}
\caption{\label{F:exp} Total energy radiated for incoherent~(top) and
coherent~(bottom) bremsstrahlung radiation in 1.5~cm thick silicon
crystal~\cite{apyan2,unel}. The solid curves represent Monte Carlo simulation
results. The experimental results are: ($\blacksquare$) - for unaligned silicon
crystal, ({\large $\blacktriangle$}) - $\psi_0 = 70 \mu rad$,
({\large $\blacktriangledown$}) -  $\psi_0 = 180 \mu rad$.}
\end{figure}

The comparison of Monte Carlo simulation data with experimental data are
given in Fig.~\ref{F:exp}. The top figure represent the simulated and measured
total radiated energy for 1.5~cm thick unaligned silicon crystal. The  
disoriented silicon crystal acts as an amorphous medium, hence we have 
incoherent bremsstrahlung spectra.

The bottom figure represents the simulated and measured CB spectra for the
two orientations of silicon crystal with respect to incident electron beam.
The electron beam was incident at an angle of $\theta_0 = 5 mrad$ to axis
$<$100$>$ for both settings of silicon crystal. For the upper curve, the
electron momentum makes an angle of $\psi_0 = 70 \mu rad$ with respect to the
$(110)$ plane and $\psi_0 = 180 \mu rad$ for lower curve.

The experimental data are taken from~\cite{apyan2,unel}. The prominent
agreement between the results of simulation and experimental data is seen.
The described Monte Carlo simulation code allows prediction of the photon
spectra, linear and circular polarizations, optimal orientation and thickness
of used crystals.

\subsection{Electromagnetic shower}
\label{shower}

The developed code is capable to track the all generations or histories of
electrons and photons in crystal. Thus, it can be used for simulation of EMS
development in single crystals oriented in CB mode. We considered EMS
initiated by the high energy electrons and photons in oriented silicon and
germanium single crystals. The incident electron and photon beams
parameters~(energy, orientation angles, angular spread) are identical with
those in NA59 experiment and discussed above (Section \ref{theory} and
Fig.~\ref{F:cros-sec}).

All calculations are carried out for the energy of initial particles
(electrons or photons) of 178.2~GeV. The simulations takes into account the
initial electron beam angular divergence in both horizontal~($48\mu rad$) and
vertical~($33\mu rad$) planes. The types of single crystals and their
orientations used in Monte Carlo simulations are presented in
Table~\ref{tab:or}. A low energy cut-off of 5~GeV was placed on the production
of all secondary particles. In this study no distinction is made between
electrons and positrons (later on simply electrons).

\begin{table}
\caption{\label{tab:or} Crystals and their orientations, radiation and
absorption lengths used in the simulations.}
                                                                                                                                                             
\begin{center}
\begin{tabular}{l l l l}
\hline
Crystal  &EMS                    &Orientation~(mrad)                        &$L_R$ or $L_A$ (cm) \\
                                                                                                                                                             
\hline
                                                                                                                                                             
Si       &Initiated by $e^-$     &$\theta_0$=30; $\psi_{0}$=0.180    &2.88~(9.36)   \\
                                                                                                                                                             
Si       &Initiated by $\gamma$  &$\theta_0$=30; $\psi_{0}$=0.6      &5.45~(12.03)   \\
                                                                                                                                                             
Ge       &Initiated by $e^-$     &$\theta_0$=30; $\psi_{0}$=0.187    &1.01~(2.30)   \\
                                                                                                                                                             
Ge       &Initiated by $\gamma$  &$\theta_0$=30; $\psi_{0}$=0.6      &1.76~(2.96)   \\
                                                                                                                                                             
\hline
\end{tabular}
\end{center}
\end{table}

Important quantities in EMS development in matter are the radiation
length ($L_R$) for charged particles and absorption length~($L_A$) for photons.
There is a weak dependence of these quantities on particle energy in amorphous
media. Practically $L_R$ and $L_A$ are constant for a given amorphous material.
There is a strong dependence of $L_R$ and $L_A$ on crystal type, its
orientation, particle energy and polarization in case of oriented single
crystals. The corresponding radiation and absorption lengths for aligned and
unaligned crystals~(in parenthesis) are given in Table~\ref{tab:or}. One can
see a large reduction of $L_R$ and $L_A$ in case of aligned crystals in
comparison with amorphous media.

\begin{figure}[htbp]
\includegraphics[scale=0.355]{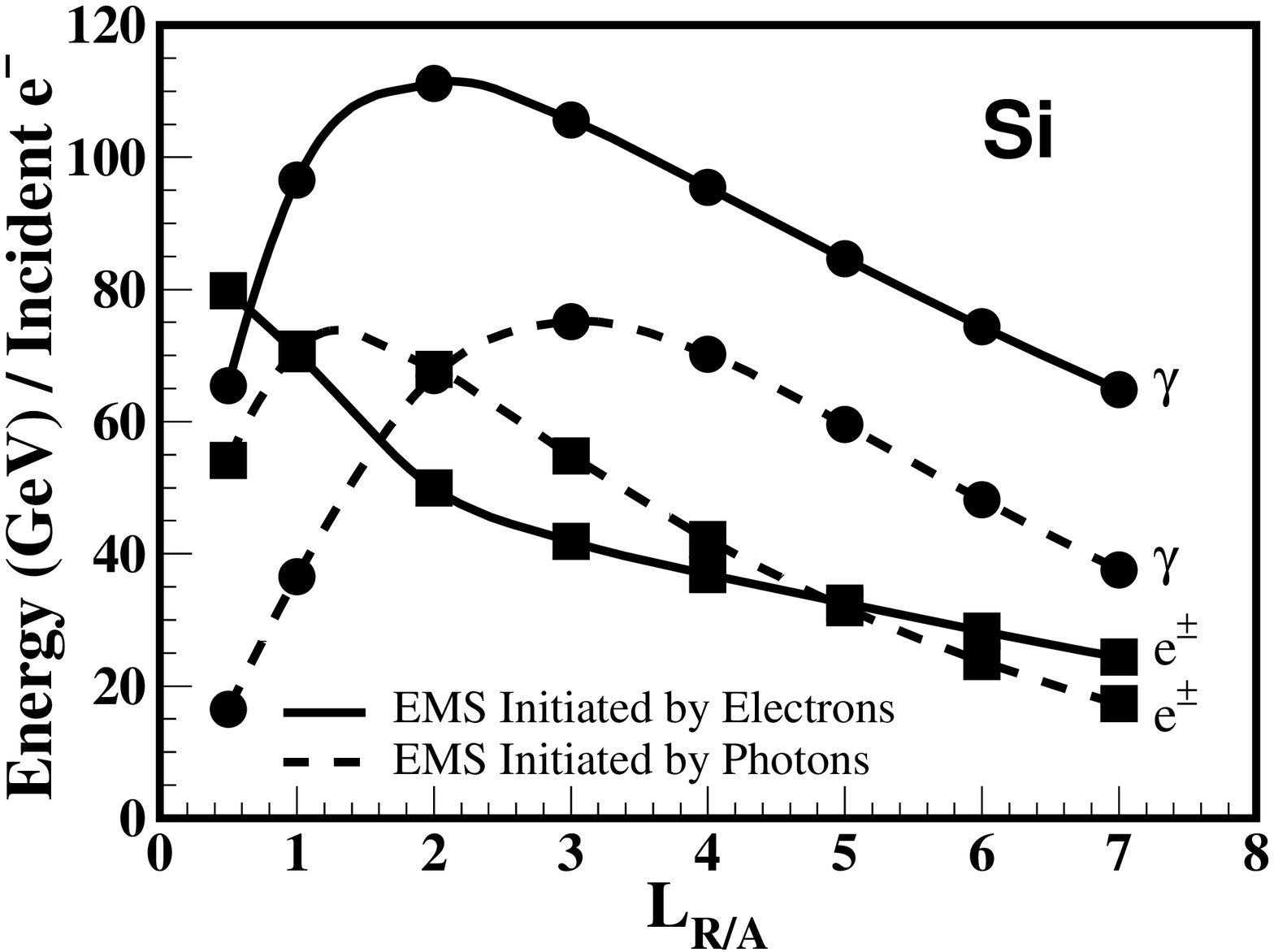} \\
\includegraphics[scale=0.355]{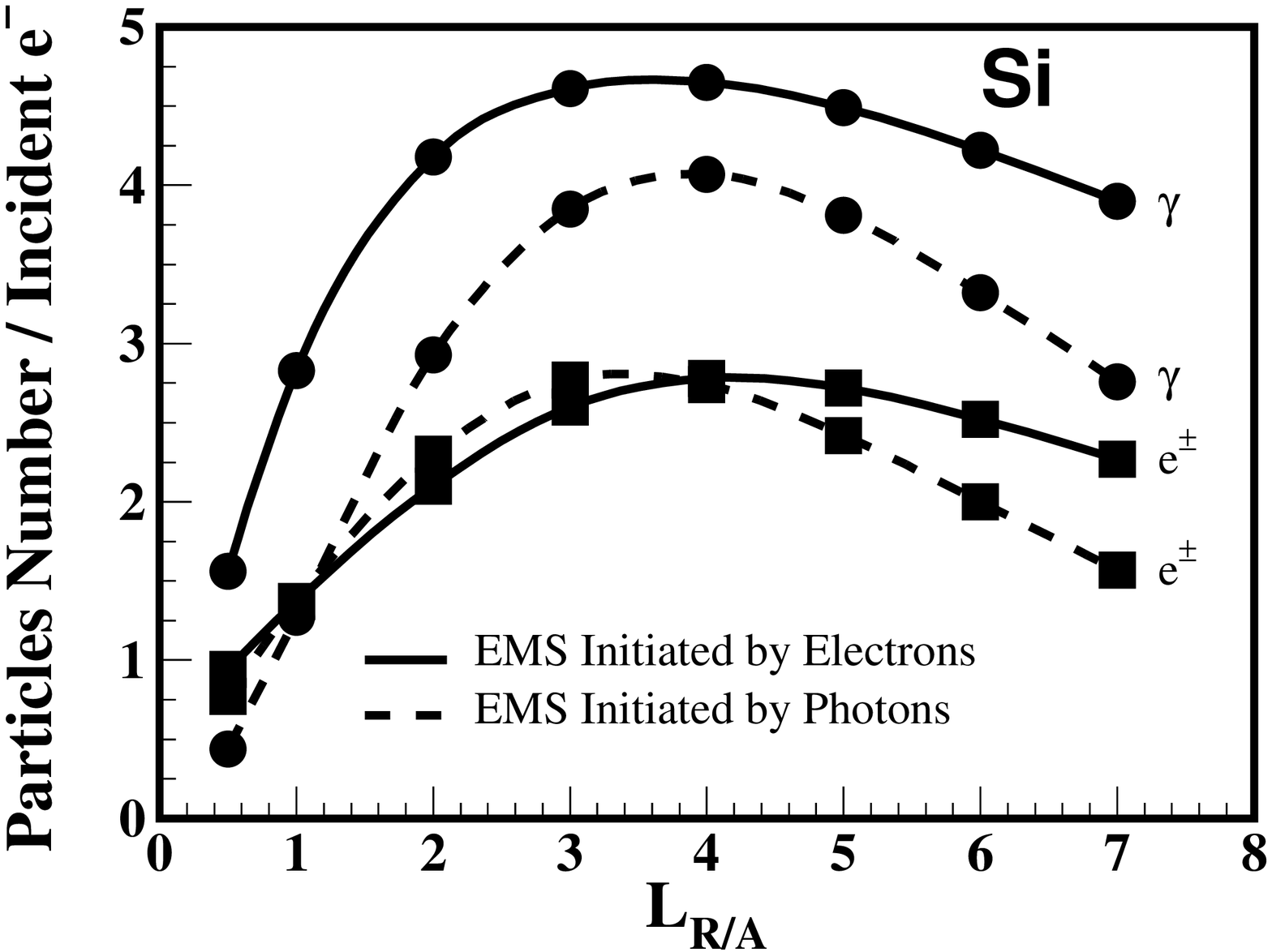}
\caption{\label{F:si} The energy~(top) and number of particles~(bottom) of EMS
as a function of thickness of an aligned silicon single crystal. $L_{R/A}$ are
the effective radiation or absorption lengths of the crystal for the EMS
initiated by electrons or photons. Solid lines represents the EMS component
initiated by electrons and dashed lines for EMS component initiated by photons.
({\Large $\bullet$}) - photon component of EMS, ($\blacksquare$) -  electron
component of EMS.}
\end{figure}

\begin{figure}[t]
\includegraphics[scale=0.355]{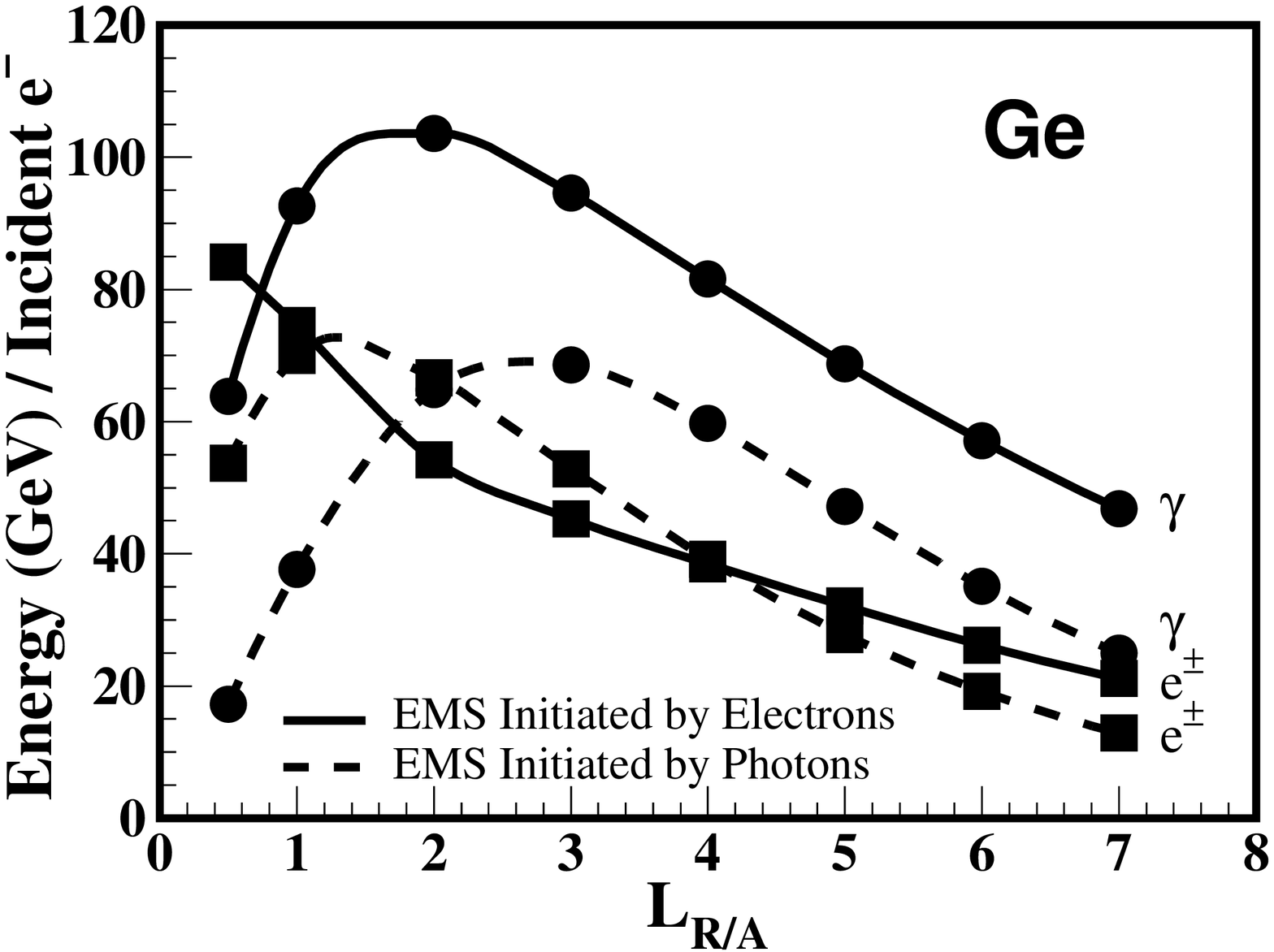} \\
\includegraphics[scale=0.355]{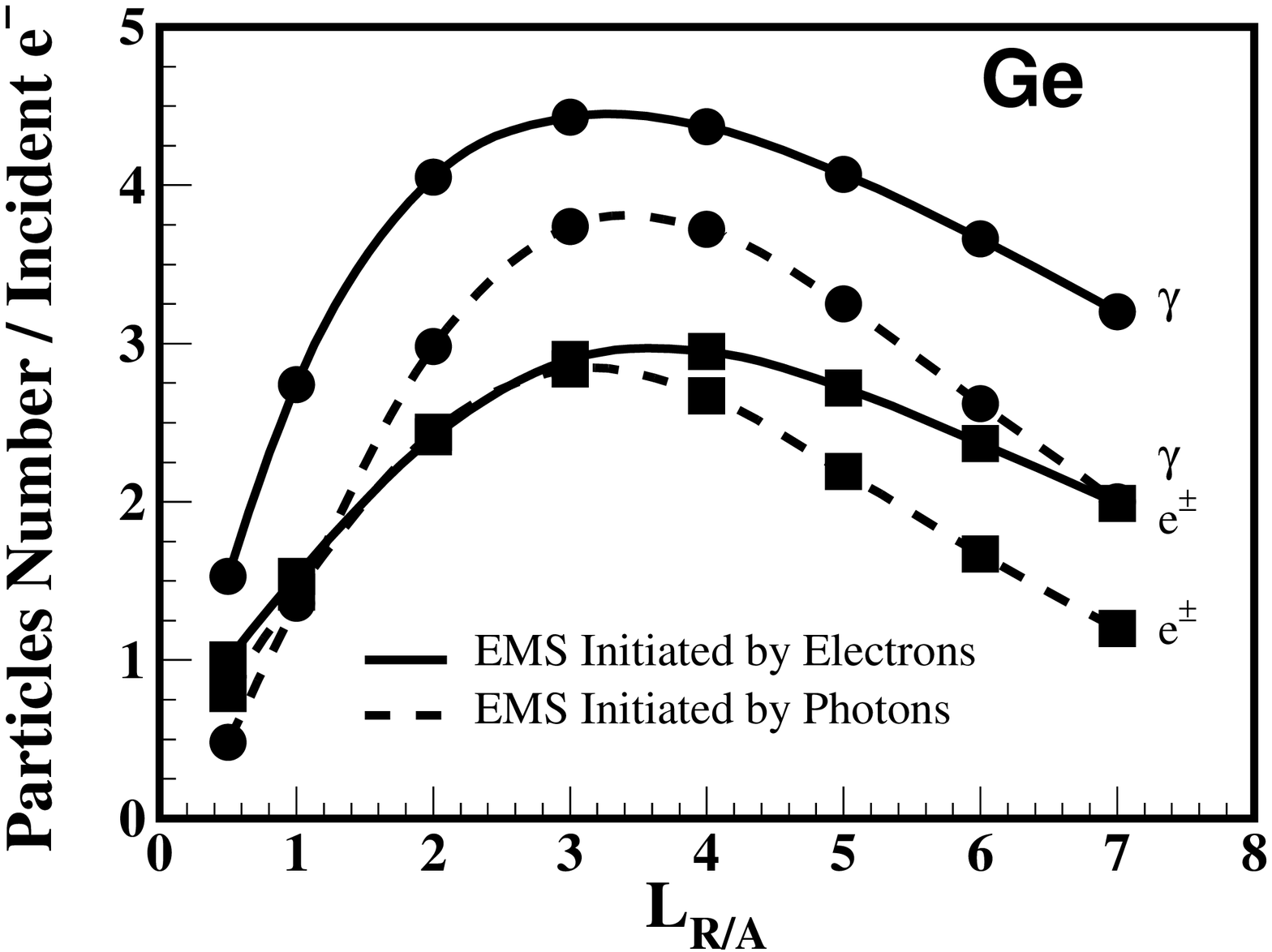}
\caption{\label{F:ge} The energy~(top) and number of particles~(bottom) of EMS
as a function of thickness of an aligned germanium single crystal. The
notations are the same as for Fig.~\ref{F:si}.}
\end{figure}

Fig.~\ref{F:si} and Fig.~\ref{F:ge} represent the Monte Carlo simulation
results of longitudinal EMS development in silicon and germanium single
crystals in energy range $5~GeV<E<175~GeV$. The energy carried by EMS is shown
on the top figures in dependence on radiation and absorption lengths. The
bottom figures show the dependence of number of particles on radiation and
absorption lengths.

The energy carried by the photon component of EMS (initiated by electrons)
reaches its maximum value at 2~$L_R$ for silicon and 1.6~$L_R$ for germanium
crystals. While the photon numbers reach the maximum at 3.5~$L_R$ for silicon
and 3.2~$L_R$ for germanium crystals. The energy carried by electrons is
smoothly decreasing for both crystals. The number of electrons reach its
maximum value at 3.5~$L_R$ for silicon crystal and for germanium crystal at
4.2~$L_R$. The photon and electron components of EMS carry $\sim$~60\% and
$\sim$~30\% of energy of initial electron, respectively, for both crystals.
The energy carried by the photon component of EMS are larger than the energy
of electrons as seen from the figures.

\begin{figure}[t]
\includegraphics[scale=0.355]{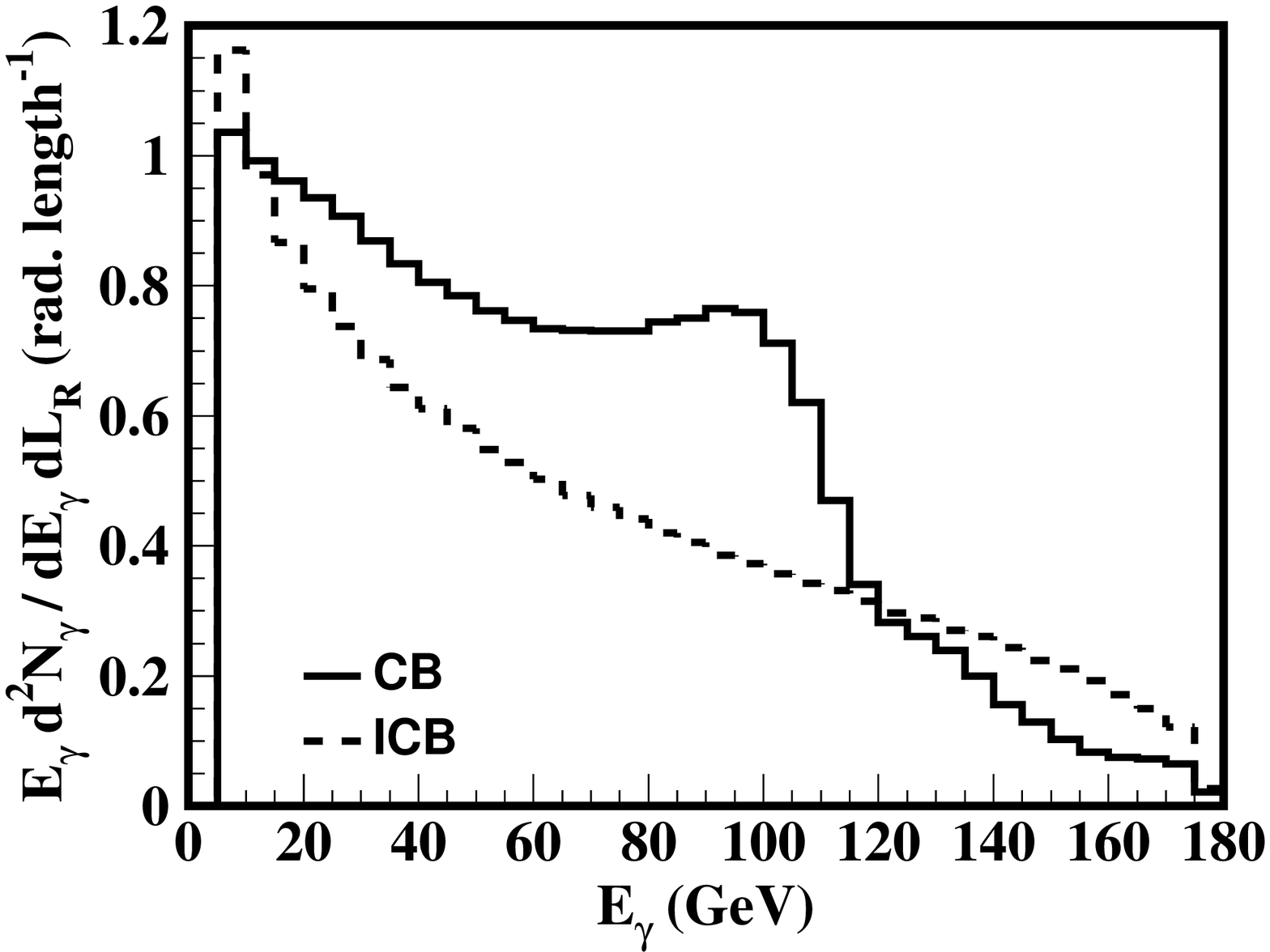} \\
\includegraphics[scale=0.355]{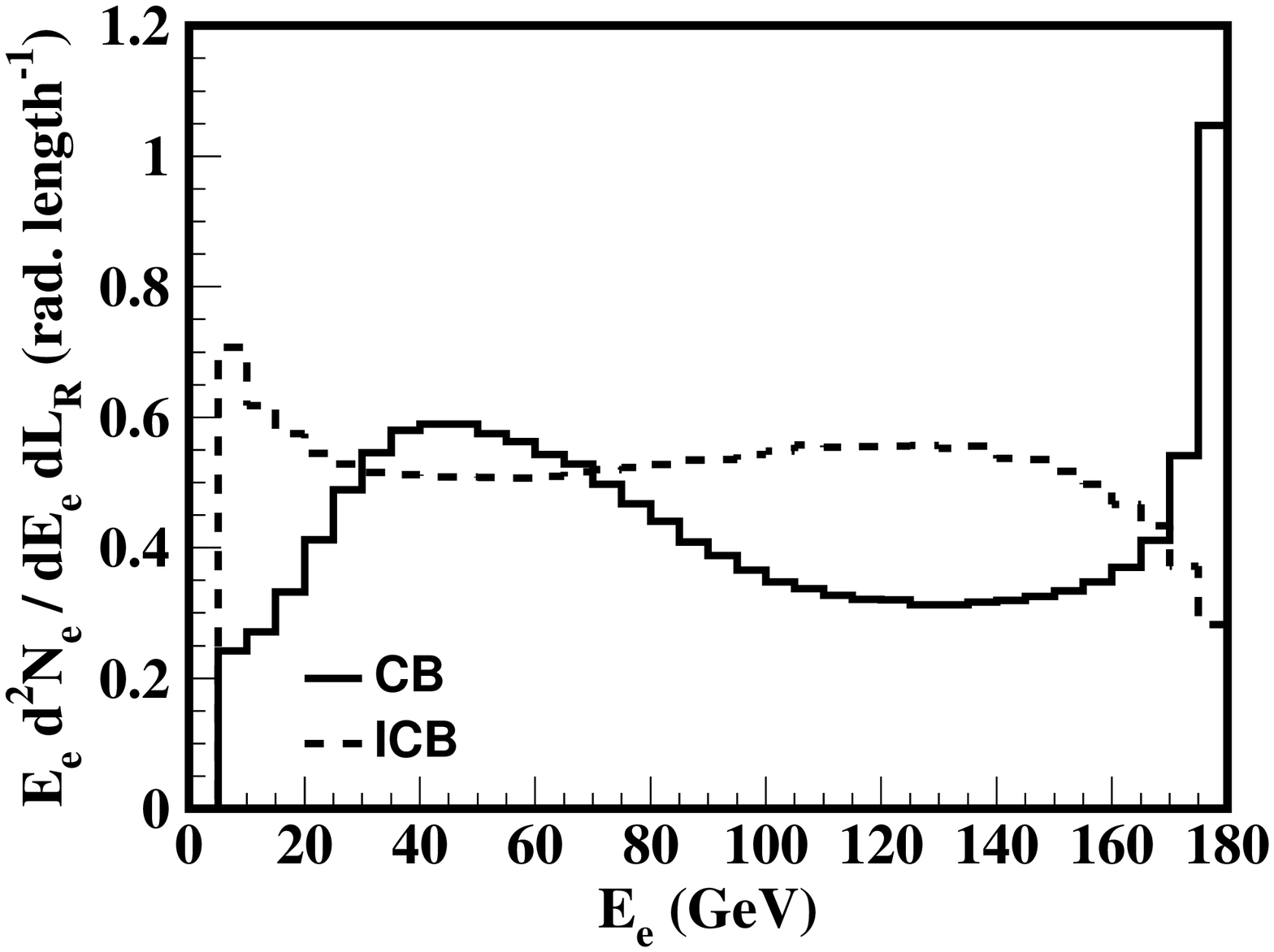}
\caption{\label{F:spectr} Simulated photon~(top) and electron~(bottom) spectra,
per unit of radiation length for a silicon crystal. The solid
curves represent the spectra from aligned crystal and the dashed curves for
unaligned crystal acted as amorphous.} 
\end{figure}

This is true for number of photons as well, which is about 2 times larger than
the number of electrons. This is explained by the CB mechanism of radiation.
Initial electrons lose large amount of energy around 100~GeV due to the CB
radiation and produce high energy photons as seen from Fig.~\ref{F:spectr}.
Simulated photon~(top) and electron~(bottom) spectra, $E d^2N/dE dL_R$, per
unit of radiation length for a silicon crystal are shown. The solid lines
represent the spectra from aligned crystal and the dashed lines for unaligned
crystal acted as an amorphous. The photon spectrum for ICB smoothly decreases 
with increasing photon energy. One can observe an expressed photon peak in the 
vicinity of 100~GeV in case of CB spectrum. The initial electron loses energy 
mainly for producing high energy photons. Thus, the huge amount of energy is 
concentrated in the high energy region of the photon spectrum. This 
concentration of energy leads to the "delay" of EMS development. For example, 
the amorphous matter or crystal oriented in channeling settings split the 
energy of electrons very effectively and EMS develops more 
intensively~\cite{baskov}.
                                                                                
The energy carried by the electron component of EMS~(initiated by photons)
reaches its maximum value at $\sim$~1.5~$L_A$ for both crystals as seen from
the Fig.~\ref{F:si} and Fig.~\ref{F:ge}. The number of electrons reaches the
maximum value at $\sim$~3.~$L_A$. The energy carried by the photon
component of EMS are approximately the same as energy of electrons at their
maximum values. But the number of photons is about 60\% larger than number of
electrons at $ L_A > 3$. The behavior of the EMS initiated by photons is
approximately the same as in the case of EMS initiated by electrons. The large
number of electron-positron pairs are produced in the vicinity of 90~GeV due
to the CPP.

\section{Conclusion}
\label{conclud}
                                                                                
The computer package presented in this paper is intended for Monte Carlo
simulation of the electron and photon beams propagation through the oriented
single crystals and amorphous media at high energies. The agreement between
the simulation results and measurements is seen to be satisfactory.
After finishing the program, the following parameters of the photons and
electrons are reported: energy, angular and spatial distributions and
polarization of passed electrons and photons. The important step in the
code is the simulation of the photon radiation and $e^{+}e^{-}$ pair
production angles, which strongly depends on the initial particles energy.
This feature of the computer code can be used for simulation of the photon
beam collimation.
                                                                                
The longitudinal behavior of electron and photon induced EMS in single
crystals oriented in CB and CPP mode in the energy range between 5~GeV and
175~GeV was investigated. The interesting results are obtained concerning EMS
development in single crystals aligned in CB or CPP mode. The crystal does
not split the energy of particles as amorphous media. It concentrates the
energy of radiation and produced $e^{+}e^{-}$ pairs in the high energy region
after the first interaction act. The crystals behave as a capacitor of energy
up to a thickness of $\sim 2 L_{R/A}$. As a result, the "delayed" EMS take
place in the crystal.

\begin{acknowledgments}
The author gratefully acknowledges Prof. M.~Velasco (Northwestern University)
for a number of very fruitful discussions.
\end{acknowledgments}

\bibliography{apyan-cosires06-O48.bib}

\end{document}